%% file: main.tex
\documentclass{article}

\usepackage{microtype}
\usepackage{graphicx}
\usepackage{booktabs} 
\usepackage{hyperref}
\usepackage{fancyhdr}
\usepackage{makecell}

\usepackage[accepted]{icml2021}
\usepackage[table]{xcolor}

\usepackage[normalem]{ulem}

\usepackage[utf8]{inputenc} 
\usepackage[T1]{fontenc}    
\usepackage{amsfonts}       
\usepackage{nicefrac}       
\usepackage{microtype}      

\usepackage{graphicx}
\usepackage{amssymb,amsmath,latexsym,dsfont}
\usepackage{tikz,pgflibraryplotmarks}
\usepackage{url}
\usepackage{amsmath} 
\newcommand{\angstrom}{\text{\normalfont\AA}}

\input{math_commands.tex}

\icmltitlerunning{Protein design and folding}

\begin{document}

\twocolumn[
\icmltitle{Mimetic Neural Networks: A unified framework for Protein Design and Folding}

\icmlsetsymbol{equal}{*}

\begin{icmlauthorlist}
\icmlauthor{Moshe Eliasof}{bgu}
\icmlauthor{Tue Boesen}{to}
\icmlauthor{Eldad Haber}{to}
\icmlauthor{Chen Keasar}{bgu}
\icmlauthor{Eran Treister}{bgu}
\end{icmlauthorlist}

\icmlaffiliation{to}{Department of EOAS, The University of British Columbia}

\icmlaffiliation{bgu}{Department of Computer Science, Ben-Gurion University of the Negev}

\icmlcorrespondingauthor{Eldad Haber}{eldadHaber@gmail.com}

\icmlkeywords{A mimetic framework for protein design and folding}

\vskip 0.3in
]

\printAffiliationsAndNotice{} 

\begin{abstract}
Recent advancements in machine learning techniques for protein folding motivate better results in its inverse problem -- protein design. In this work we introduce a new graph mimetic neural network, MimNet, and show that it is possible to build a reversible architecture that solves the structure and design problems in tandem, allowing to improve protein design when the structure is better estimated. We use the ProteinNet data set and show that the state of the art results in protein design can be improved, given recent architectures for protein folding. 
\end{abstract}

\section{Introduction}

Protein folding has been an open challenge in science for many years \citep{Finkelstein2004,Oidziej7547, RoseEtAl2006,AlQuraishi2019}. 
The goal of protein folding is to predict the 3D structure of a peptide chain given its amino acids (residues) composition. 
Traditional methods for folding were based on physical understanding of the interaction potentials. However, they require considerable computational resources and often converge to a local minima \citep{NedwidekHecht1997}. In recent years, techniques based on machine learning, and in particular on deep neural networks, have been proposed for the solution of the problem,  \citep{AlQuraishi2019,drori2019accurate,Xu16856,alphaFold} showing major improvement in protein structure estimation. Such methods utilize advances in machine learning and network architectures and perhaps more importantly, large protein data sets and advanced, nontrivial pre-processing techniques. These data sets are used in order to find homologous proteins by computing  Multiple Sequence Alignment (MSA). The MSA is then used to compute first order statistics in terms of Position-Specific Score Matrices (PSSM) and second order statistics in terms of the covariance matrices of the homologous proteins. Since the second order statistics  can yield an approximation to contact maps \citep{Vassura2008}, such tools can aid in hinting about the structure and guide the training of the network.
 Nonetheless, using second order statistics such as covariance matrices and contact maps requires non-trivial computations that are time consuming, in particular searching through large data-bases computing MSA's and their approximate covariances. It has therefore been proposed in \citep{AlQuraishi2019} to use first order statistics in the form of PSSM for the task in hand. While such approach does not yield state of the art results for the protein folding problem, it is significantly faster to obtain and train. For the purpose of protein design, such first order statistics may be sufficient, as our experimental results suggest in Sec. \ref{sec:designExperiment}. For the protein folding problem, we demonstrate that by utilizing Graph Convolution Networks (GCNs), it is possible to improve accuracy compared to state of the art models which use only first order data, in section \ref{sec:foldingExperiment}.

A closely related problem to the folding that did not get as much attention is the protein design \citep{Richardson1989,Yennamalli2019,Basanta22135}.
In the latter we assume to have a known shape of a protein and the goal is to find a plausible underlying sequence. The problem is often referred to as the inverse problem of protein folding. In recent years, deep learning techniques have been used for the design problem \citep{Wang2018Design,GAO2020,Strokach2020, Xu2020design} with promising success. Nevertheless, the treatment of the two problems has been disjoint. Therefore, techniques in protein design do not leverage advances in folding and vice versa.
It is well known that the design problem does not have a unique solution. Indeed, there may be more than one sequence that yields the same or at least very similar structure \citep{Vassura2008}. In fact, many homologous proteins share similar if not identical (up to measurement errors) structures.
Therefore, judging the success of the prediction by looking at a single sequence can be misleading. To this end, we propose to approach the problem by predicting a family of protein sequences with similar structures. Namely, given a protein structure (coordinates), we aim to predict its PSSM, instead of finding a unique sequence that describes it. Notably, enriching a protein sequence by family-consensus residues often leads to better structural stability, a strategy know as back-to-consensus \citep{BERSHTEIN2008, Chandler2020}. 

{\bf Previous and related work} \\
Recently, the problem of protein folding drew large attention with the recent summary of the CASP 14 competition \cite{Liu2021.01.28.428706}. However, using first order information only (that suffice for protein design) received notably less attention. Related works to ours can be found in \cite{Li-etal2017, Gao-et-al2018,AlQuraishi2019, TORRISI20201301}.
As we show in our numerical results in Sec. \ref{sec:experiments}, our approach yields superior results given the first order data. 

Employing deep learning for protein design task is a relatively new idea. A similar approach to ours was recently presented in \cite{Strokach2020} where graph methods were proposed for protein design, reporting
promising results by treating the problem as a graph node classification problem, surpassing other de-novo design codes. 
While \cite{Strokach2020} and our method share some similarities, ours is largely different as we use reversible architectures which offer numerous advantages, discussed in the following. As we show in Sec. \ref{sec:designExperiment}, our approach obtains better results on a large data-set derived from the Protein Data Bank (PDB).

{\bf Main Contribution} \\
The main part of this work is the introduction of a framework that unifies the treatment of protein folding and design. Our framework mimics the physics of folding a protein using a neural network. Hence, we coin the term \textit{Mimetic Deep Neural Networks} (MimNet), which we apply to graphs, describing protein structures.  While our work focuses on protein folding and design, the proposed network can be applied with any node or edge data that is available, thus it suits using both first or second order statistics.

The main idea is to generate a reversible transformation from the structure to the sequence and vice versa by using reversible architectures. These networks allow us to jointly train the folding and the design 
problems, utilizing both the sequence and the structure of the protein {\em simultaneously}. We explore a family of neural networks that are designed to do just that. These are neural network architectures that are inspired by Hamiltonian dynamics and hyperbolic differential equations \citep{chang2018reversible, ruthotto2019deep}. Such networks can propagate forward and backward and, they can utilize any type of layer, from structured to graph convolution or attention, harnessing recent advances in the understanding of protein folding architectures. In this paper, we particularly explore the use of Multiscale Graph Convolutional Networks (Multiscale GCNs), which are graph-based deep learning methods \citep{gao2019graph,wang2019dynamic, eliasof2020diffgcn}. Such networks can efficiently mimic the pairwise interaction of the potential in a three dimensional physical system.

Reversible networks are bidirectional, and therefore it is natural to train them for both folding and design simultaneously, effectively doubling the amount of the data with respect to the network parameters.
Another important advantage of such a network is its memory footprint. Since the network is reversible, it is possible to train an {\em arbitrarily long} network without storing the activation, at the cost of double the computation of the backward pass \cite{chang2018reversible}. This enables the use of very deep networks that are impossible to use otherwise.

The physical folding process can be described by a second order differential equation derived from Hamiltonian dynamics. Reversible architectures that are inspired by Hamiltonian dynamics can be used to simulate this process. One can therefore claim that such a mimetic network is more faithful to the physics of the protein folding problem compared to a standard deep network like a ResNet \citep{he2016deep}.

The rest of the paper is organized as follows. In Sec.~\ref{sec:method} we discuss the problem and introduce the key mathematical ideas which constitute the building blocks of our network. In particular, we discuss multiscale reversible networks and different types of graph convolution techniques that are used to solve the problem. We then define our MimNet and its objective functions. In Sec.~\ref{sec:experiments} we perform numerical experiments with data obtained from ProteinNet \citep{ProteinNet}. ProteinNet is a publicly available data set that contains both sequences and PSSMs and thus allows for the training of a folding network with first order statistics as done in \citep{AlQuraishi2019}. The size of the data set, its structure and the division into training validating and testing, which were carefully selected, allows one to 
rigorously test the design problem as well.
Finally, in Sec.~\ref{sec4} we discuss the results and summarize the paper.

\section{Methods}
\label{sec:method}

Before discussing the particular network and architecture, we define the data and the functions of folding and design problems. 
Specifically, assume that $\bfS \in {\gS}$ is a $20 \times n$ matrix that represents a protein sequence of $n$ amino acids. Let  $\bfS^+ \in {\gS}^+$ be additional data that is related to the sequence such as PSSM and possibly covariance information derived from MSAs. Also, let $\bfX \in {\gX}$ be a $3\times n$ matrix that represents the protein structure (coordinates). We define the 
mapping $F:{\gS}\times {\gS}^+ \rightarrow {\gX}$ as the folding mapping. 
This mapping takes the information in $\bfS$ and $\bfS^+$ and maps it into the estimated coordinates  $\hat{\bfX}$ that reveal the structure of the protein. Throughout the paper we denote $F(\bfS)$ instead of $F(\bfS, \bfS^+)$ for brevity. Consider now the opposite mapping from the space $\gX$ to the space $\gS \times {\gS}^+$. We denote this mapping as $F^{\dagger}:\gX \rightarrow \gS \times {\gS}^+$ and it can be thought of as some psedu-inverse of the mapping $F$. 
These mappings can be learnt separately and independently as has been done so far. However, since $F$ and $F^{\dagger}$ are closely related, it is tempting to jointly learn them, utilizing both the sequence, its attributes, as well as the structure of the protein {\em in tandem}. 

We now review the concept of a mimetic deep neural network, that is, a neural network that mimics the physics of the dynamics of the folding process. To this end, a deep network can be thought of as a time discretization of a differential equation \citep{chen2019neural,ruthotto2019deep}. According to this interpretation, each layer represents the state of the system at some particular pseudo-time.  The mimetic properties are first discussed in pseudo-time, namely, how the network propagates from one layer to the other. The second mimetic property considers the spatial domain, meaning, how a particular residue in the protein interacts with another residue. These properties are put together to generate a mimetic deep neural network that imitates molecular dynamics simulations using network architectures that are derived by discretized differential operators in time and space \citep{ruthotto2019deep, eliasof2020diffgcn}. The treatment in both space and time are put together within a network optimization procedure to train the system and yield a network that can solve both the folding and the design problems.

\subsection{Reversible Networks and Dynamical Systems}
\label{sec:reversible_nets}
In this subsection we show how to build a mimetic network in time by using reversible dynamics.
Reversible systems play a major role in physics for applications that range from Hamiltonian dynamics to
wave equations. Broadly speaking, a reversible system is one that can propagate forward in time without information loss and therefore, can propagate backwards in time. Simple physical examples are a pendulum 
or a wave. These systems (in their idealized form) do not change their entropy, and therefore allow for forward or backward integration in time. Typical molecular dynamics is solved using reversible methods \citep{Saitou1987} (that is, integrating Hamiltonian dynamics) and therefore, it is natural to explore neural network architectures with similar properties.

To be more specific, given the input for the folding task $[\bfS, \bfS^+]$ (e.g., the concatenation of the one hot encoding sequence design and PSSM matrices) we first apply 
\begin{eqnarray}
\label{eq:openLayer}
\bfY_0 = q([\bfS, \bfS^+], \bftheta_e)
\end{eqnarray}
where $\bfY_0$
contains $n_f$ channels of $n$-length sequence features, embedded by the transformation $q(\cdot,\cdot)$, parameterized by the weights $\bftheta_e$. This layer transforms the input to the latent space of the network. Here we use a 1D convolution for $q$, but other transformations may also be suitable. 
 
 The initial state $\bfY_0$ and its velocity vector $\bfV_0$ are then pushed forward by a deep residual neural network.
In particular, we consider a network with the following structure \begin{subequations}
\label{hamnetgen}
\begin{eqnarray}
\bfV_{j+1} &=& \bfV_j + h \cdot f(\bfY_j, \bftheta_j) \\
\bfY_{j+1} &=& \bfY_j + h \cdot g(\bfV_{j+1}, \bftheta_j), 
\end{eqnarray}
\end{subequations}
where $j=0,\ldots,T$ is the layer index. 
$h$ is a parameter that represents a time step size and $\bftheta_j$
are learnt parameters that characterize the $j$-th layer. The system in \eqref{hamnetgen} can be interpreted as a Verlet type discretization of a dynamical system with learnable forces that are the gradients of some potential function.
A particular case of such dynamics is obtained by setting $g = Id$ (the identity transformation) yielding the second
order dynamics 
\begin{eqnarray}
\label{hypernet}
\bfY_{j+1} &=& 2\bfY_j  - \bfY_{j-1} + h^2 f(\bfY_j,\bftheta_j). 
\end{eqnarray}
This scheme is clearly reversible, regardless of the choice $f$ (which we discuss Sec. \ref{sec:GraphUnet} ), since we can express $\bfY_{j-1}$ as a function of $\bfY_j$ and $\bfY_{j+1}$. 
The propagation forward (and backward) is not complete without defining the boundary conditions $\bfY_{-1}$ and $\bfY_{T+1}$. Here we arbitrarily choose $\bfY_{-1} = \bfY_0$ and $\bfY_{T-1} = \bfY_{T}$, that is, initializing the network with zero velocity, i.e., $\bfV_0 = 0$. An illustration of the dynamics is plotted in Fig.~\ref{fig1}.

\input{Fig1.tex}

Given the final state of the system $\bfY_T$, we predict the coordinates $\bfX$ by projecting $\bfY_T$ onto a 3 dimensional space 
\begin{eqnarray}
\label{eq:embedcoords}
\hat\bfX = q^+(\bfY_{T},\bftheta_f),
\end{eqnarray}
where $\hat\bfX$ are the predicted coordinates. The transformation $q^+(\cdot,\cdot)$ can be realized by a neural network, and we choose it to be a learnable projection matrix of size $n_f \times 3$ such that the final feature maps are projected to 3D coordinates.

The layer in \eqref{eq:embedcoords} may also contains some additional constraints. In particular, we may demand that
$$ |\hat\bfX_i - \hat\bfX_{i-1}| = c, $$
constraining the distance between every two residues to $c=3.8 \angstrom$. We have found that when the data is noisy implementing this constraint is needed in order to obtain physical results (see Sec. \ref{sec:foldingloss}).


In the forward pass, described above, the folding problem was solved, where we march from the protein design attributes (as in Sec. \ref{sec:reversible_nets}) to its coordinates. In the backward pass, we solve the design problem, where our goal is to predict the sequence given its coordinates. We start the backward pass by embedding the coordinates into the network feature space, i.e.,
\begin{equation}
\bfY_T = (q^+)^*(\bfX,\bftheta_f),
\end{equation}
where $(q^+)^*$ is the adjoint of the transformation $q^+$. We then march backwards, replacing the entries of $\bfY_{j+1}$ and $\bfY_{j-1}$ in \eqref{hypernet} and finally, using the adjoint of $q$ to propagate from $\bfY_0$ to the sequence space
\begin{eqnarray}
\label{eq:coordToDes}
[\hat\bfS,\hat\bfS^{+}] = q^{*}(\bfY_{0},\bftheta_e),
\end{eqnarray}
where $[\hat\bfS,\hat\bfS^{+}]$ are the predicted protein design attributes.
These forward and backward passes couple the design and the folding tasks together into a single network that, similarly to the physical dynamics, can be integrated (in time) from sequence to coordinates and backwards from coordinates to a sequence. 

\subsection{Graph Convolutional Networks}

Sec. \ref{sec:reversible_nets} considers the propagation of the network from its initial condition (a sequence) to its final one (3D structure) and vice versa. The discussion was agnostic to the choice of the function $f(\cdot,\cdot)$ in \eqref{hypernet} that realizes the network in hand. In this section we review the concept of a graph network and discuss its computation.

The idea behind a graph based method is rooted in the physics of the problem. Energy based simulations
can be thought of as pairwise interactions on a graph based on the $L_2$ distance between the residues. Indeed, as the distance between residues is smaller, the interaction between them is stronger.
This motivates us to use machine learning techniques that mimic this property. As the dynamical system is evolving, the interaction between pairs of close residues is significantly larger compared to far ones.

One of the most successful techniques for image and speech processing is Convolution Neural Networks (CNN) \citep{krizhevsky2012imagenet,Goodfellow-et-al-2016}. The method relays on the structured grids on which sequences and images are defined. That is, every element has neighbouring elements in a structured manner. In recent years, similar ideas were extended to more complex geometries and manifolds, 
which can be naturally represented by a graph
\citep{ranjan2018generating,hanocka2019meshcnn,wang2019dynamic}. The main idea is to replace the structured convolution with a graph based convolution. That is, rather than convolving each location with its near neighbours defined by the sequence, define the distance between each location based on the graph node or edges features, and then convolve the residues that are close on the graph. 

To be more specific, we let $\bfY_j$ be the state at the $j$th layer. Then, we define a graph convolution block as follows:
\begin{eqnarray}
\label{eq:gcn}
f(\bfY_j) &=& -\mathcal{C}^{*}\left(\theta_{j},\sigma\left(\mathcal{C}(\theta_{j},\bfY_j\right)\right))
\end{eqnarray}
where $\mathcal{C}(\theta_{j},\cdot)$ is the graph convolution operator with its learned associated weights $\theta_j$. This operator spatially resemble a discrete differential operator, e.g., a mass term, a graph Laplacian, or an edge gradient \citep{eliasof2020diffgcn}. $\sigma(\cdot)$ is the ReLU activation function. The operator $\mathcal{C}^*$ is the adjoint operator of $\mathcal{C}$ (like a transposed convolution), applied using the same weights $\theta_{j}$. This way, assuming that $\sigma$ is a monotonically non-decreasing function that either zeroes its input or preserves its sign, we get a symmetric and positive semi-definite operator. We use the negative sign in front of the layer such that
the operator $f(\cdot)$ is negative, which is important if we are to generate a stable dynamics---see \citep{ruthotto2019deep} for details and analysis.

 
Many graph based networks employ a graph convolution with fixed connectivity \citep{ranjan2018generating, bouritsas2019neural}. This is reasonable if the final topology is known. However, for protein folding we start with an unknown structure and it is evolving (learnt) from the data. Therefore, rather than using a fixed graph for the network we let the graph evolve throughout network. We thus recompute a weighted graph Laplacian at each layer, or, for computational saving, every $\frac{T}{3}$ U-net layers. To this end, we compute the weighted distance matrix between each two residues
 \begin{eqnarray}
 \label{eq:weightByDistance}
\bfW_j = \exp\left(-\alpha^{-1} \bfD(\bfY_j)\right),
\end{eqnarray}
where $\alpha$ is a scaling parameter (we set $\alpha = 10$) and $\bfD$ is the $L_2$ distance between each two residues 
\begin{equation}
\label{distmat}
\bfD(\bfY) = \left(\bfY^2 \bfone \bfone^{\top} + \bfone \bfone^{\top} \bfY^2 - 2 \bfY^{\top} \bfY\right).
\end{equation}
The vector $\bfone$ is a vector of ones of  appropriate size.  Using the distance matrix we define the graph Laplacian as
\begin{eqnarray}
\label{gL}
\bfL_j = {\rm diag} (\bfD_j \bfone) - \bfD_j.
\end{eqnarray}
The approach of dynamically updating the connectivity of the graph was also suggested in \citep{wang2019dynamic}. Our strategy differs in that instead of picking $k$ nearest neighbors to be equally weighted, regardless of their distances, we use a weighted and fully connected graph.
 That is, our graph Laplacian is a dense matrix. This is reasonable since the typical size of a protein is less than $1,000$ residues with a mean size of $350$, and similarly to various physical applications, every two residues interact according to their distance \citep{NedwidekHecht1997}. Further, the weighting of the edges makes the graph Laplacian continuously differentiable with respect to the network, which aids its training.

\begin{figure*}
    \centering
    \includegraphics[width=0.92\textwidth]{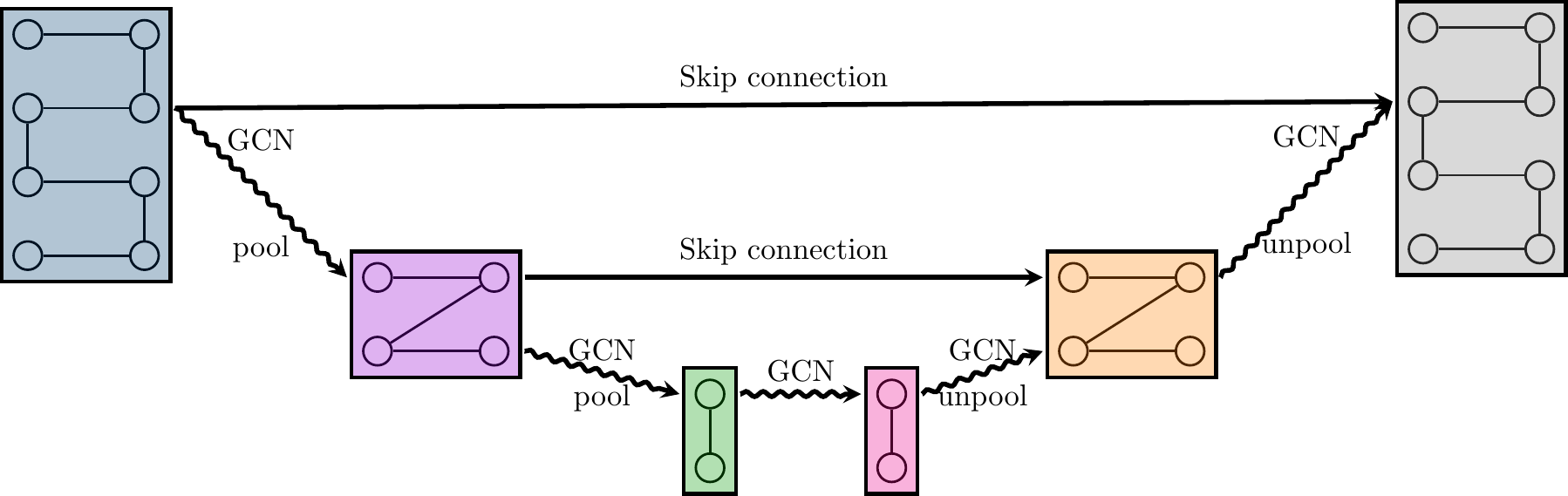}
    \caption{A graph U-net. GCN is defined in \eqref{eq:gcn_relax}. Pool and unpool denote graph coarsening and refinement, respectively. Skip connection denotes a summation of the respective feature maps.
    }
    \label{fig:unet}
\end{figure*}

\subsection{Multiscale Graph Networks}\label{sec:GraphUnet}

The limitation of graph based networks, similar to other convolution methods is that they generate strong local interactions only. Hence, spatially-distant connections may suffer from weak interactions (due to small weights), and information will be spread slowly within the network - requiring more layers to compensate for. An elegant way to have long interactions and pass information between far-away parts of the graph is to consider a multiscale framework.

To this end, instead of a standard graph convolution $\mathcal{C}$ in \eqref{eq:gcn}, we use a multiscale mechanism that is similar to a U-net \citep{ronneberger2015unet,shah2018stacked}, where coarse scale approximations of the protein are composed. In particular, in the multiscale version of \eqref{eq:gcn} we choose $\mathcal{C}$ in  to be the encoder part of a U-net, and the operator $\mathcal{C}^*$ is the transposed operation, that has a decoder structure (parameterized by the same weights). Together, they form a symmetric graph U-net. The reversibility of the networks remains, since \eqref{hypernet} is reversible for every $f$, and in particular for our symmetric U-net.
  
Our graph U-net is comprised of $n_{Levels}$ graph scales. At each level we perform a GCN block where we use both the graph and sequence neighbors in our convolutions:
\begin{eqnarray}
\label{eq:gcn_relax}
\bfY_{j+1} &=& \omega_j\bfY_j + \sigma(\mathcal{N}(\bfK_j(\bfY_j + \bfY_j\bfL_j))). 
\end{eqnarray}
where $\bfK_j$ is a 1D convolution with kernel of size 9, connecting nodes on the protein sequence, and $\bfL_j$ is the graph Laplacian operator from \eqref{gL}.
 $\mathcal{N}$ is the instance normalization layer, and $\sigma$ is the ReLU activation. $\omega_j$ equals $1$ when graph coarsening is not applied, and $0$ otherwise. On the coarsest level of the U-net, we perform 2 convolution steps like \eqref{eq:gcn_relax}.
At each level, the graph differential operator $\bfL_j$ is re-computed on the coarse graph allowing for simple and inexpensive computations between scales.
In addition, since the protein has a simple
linear underlying chain, we use linear coarsening, implemented by simple average 1D pooling---this is illustrated in Fig.~\ref{fig2}. \input{Fig2.tex} 
In the decoder part of the U-net we apply the transposed operators, and to refine our graph (unpooling) we use a linear interpolation along the chain. To propagate information between matching levels we 
add long skip-connections after each convolution, for a stable training scheme (see Fig. \ref{fig:unet})

\subsection{The MimNet Architecture}
\label{sec:buildingMIMnet}

Combining our building blocks together, we now define our bi-directional mimetic architecture, called MimNet. The network consists of three main components - the opening embedding layer, a stacked graph U-net modules, and a closing embedding layer. 

At the start and end of MimNet we use the embedding layers \eqref{eq:openLayer} and \eqref{eq:embedcoords}, both of which are implemented using a simple $1 \times 1$ convolution of appropriate sizes ($n=40$). At the core of our network we employ a series of $T$ graph U-nets modules. Each graph U-net is defined according to section \ref{sec:GraphUnet}, and all of them are of identical dimensions. That is, each has $n_f$ channels of equal dimensions on the finest level. 



\subsection{Training MimNet}

Our MimNet allows us to build the physics of the problem into the neural network. This needs to be followed by a thoroughly thought training process. 
In particular, care needs be taken when choosing the appropriate problem to minimize and the appropriate choice of loss functions and regularization. We now discuss these choices for our training.

\subsubsection{The optimization problem}

Since we have a bidirectional network, we use both directions to train the network. We define the objective function
\begin{eqnarray}
\label{objfun}
{\gJ}(\bftheta) &=& {\frac 1N}\sum_j {\ell}_{fold}(F(\bfS_j,\bftheta), \bfX_j)  \\ &&+{\frac 1N}\sum_j\ell_{design}(F^{\dagger}(\bfX_j,\bftheta), \bfS_j) + \beta R(\bftheta). \nonumber
\end{eqnarray} 
Here $\bftheta$ are all the parameters of the network, $F$ is the forward network from sequence to coordinates  and $F^{\dagger}$ is the backward mapping from coordinates to sequence. The loss functions ${\ell}_{fold}$ and ${\ell}_{design}$ are chosen to measure the discrepancy between the estimated and true coordinates and between the  predicted and true sequence design, respectively. The choice of these functions is to be discussed next. Finally, $R(\cdot)$ is a regularization term that ensures stability of the network and is described below.

\subsubsection{Loss Function for the Design Problem}
\label{sec:lossDesignTask}
The loss function for the design problem is rather straight forward. At every residue location we attempt to recover the individual residue out of $20$ possible ones. Noting
\begin{equation} 
\hat\bfS = F^{\dagger}(\bfX_j,\bftheta) 
\end{equation}
we interpret $\hat\bfS$ as a matrix that its $ij$ entry is the probability of the $j$th residue to be of residue type $i$. Therefore, it is straight forward to use the cross entropy as a metric,  setting
\begin{equation}
 {\ell}_{design} = \sum \bfS \odot \log(F^\dagger(\bfX,\bftheta)). 
\end{equation}
Note that the network output is exactly the definition of the PSSM, therefore, 
if one assumes that the PSSM of the sequence is available, then, it is possible to use the KL divergence between the computed and observed PSSMs as a distance metric.  This is because the PSSM represents the true probability of the $j$th residue in the sequence to be of type $i$. 
While this strategy is always possible during training, using it for inference can be difficult. Indeed, no mapping known to us is given from PSSM to a particular residue. However, having a PSSM as an answer may allow for greater flexibility when designing a protein, since there is not necessarily a unique answer to the design process, and the PSSM represents this ambiguity. One can always use the maximum probability of the PSSM for the design a particular protein. Although there is no guarantee, this often leads to better structural stability \citep{BERSHTEIN2008, Chandler2020}.

\subsubsection{Loss Function for the Folding Problem}
\label{sec:foldingloss}
We turn our attention for the loss function for the folding problem. Clearly, one cannot simply compare the coordinates obtained by the network, $F(\bfS)$ to the observed coordinates of the sequence, as they are invariant with respect to rotation and translation. Similar to the work \citep{AlQuraishi2019} one can compare the distance matrices obtained from the coordinates.
Let $\bfD_s(\bfX) = \sqrt{\bfD(\bfX)}$ be the pairwise distance matrix in~\eqref{distmat}. 
The distance matrix is invariant to rotations and translations. Thus, it is possible to compare the
distances obtained from the true coordinates, $\bfD_s(\bfX)$ to the distances of the predicted coordinates
$\bfD_s(F^{\dagger}(\bfS))$ by their dRMSD in \eqref{eq:drmsd}.


In an average protein, residue distances typically range from blue dozens of Angstroms to a few Angstroms. Minimizing the $L_2$ distance is therefore focused on the large scale structure of the protein and can neglect the small scale structures as they contribute remarkably less to the loss.
Also, it is well known that the distance between 1-hop (immediate) neighboring residues is smaller than $3.8 \pm 0.04 \angstrom$.
This motivated previous works to use a threshold value and ignore distances larger than that threshold.
For example, in AlphaFold \citep{alphaFold} a distance of $22 \angstrom$ was used the cutoff. Here we used a slightly more conservative value of $7$ residues which translates into 
$7\times 3.8 \angstrom  = 26.6 \angstrom$.

Another aspect of the particular PDB data is that it can be noisy. As a result, some of the distances between residues are not physical. In particular, distances that are $\ell$ apart cannot have distance that is larger than $\ell \times 3.8 \angstrom$. Unfortunately, the data presents many such pairs. Additionally, some of the residues are missing information (whether their sequence or coordinates).
We therefore do not consider such entries of the data, by masking them during training and inference.
To summarize, the loss can be expressed as
\begin{eqnarray}
\label{eq:drmsd}
{\ell}_{fold} = \sqrt{{\frac 1 {n_M}}\left\| \bfM \odot \left(\bfD(F(\bfS,\bftheta)) -  \bfD(\bfX) \right) \right\|_{F}^2}
\end{eqnarray}
where $\bfM$ is a masking matrix that masks the part of the data that is not-physical or missing and $n_M$ is the number of non-zeros in the matrix. 

\subsubsection{Regularization}

The last component in our optimization scheme is the regularization, $R(\bftheta)$. We rewrite $\bftheta = [\bftheta_0, \ldots, \bftheta_L]$ where $\bftheta_j$ are the parameters used for the $j$-th layer.
Then, stability for the dynamical system is obtained if the total variation of its parameters is small \citep{ruthotto2019deep}. Thus we choose the following regularization function
\begin{eqnarray}
\label{regfun}
R(\bftheta)  = \sum_j |\bftheta_{j+1} - \bftheta_j|_1
\end{eqnarray}
Note that we do not use the standard Tikhonov regularization (so called weight decay) on the weights as they do not guarantee smoothness in time which is crucial for reversible networks and integration in time \citep{celledoni2020structure}.                         


\section{Numerical Experiments}
\label{sec:experiments}
We verify our method by performing three sets of experiments - protein folding, design, and an ablation study to quantify the contribution of the reversible learning.

\subsection{Dataset and Settings}
\paragraph{Dataset}
For the experiments we used the data set supplied by the ProteinNet \citep{ProteinNet}. The data contains proteins 
processed from the PDB data set, and is organized to hold training, validation and testing splits specifically for CASP $7-12$. For instance, the ProteinNet on CASP $11$ data contains 42,338 proteins that are less than 1000 residue long for training, 224 proteins for validation and 81 test proteins. The data set was used in \citep{AlQuraishi2019} and more recently in \citep{drori2019accurate}. We use the $90\%$ thinning version of the data, as reported in \cite{AlQuraishi2019}. While the first order statistics is  available, second order statistics cannot be downloaded freely and requires complex and expensive pre-processing. We therefore use only first order statistics in this work and compare it to other recent methods that uses identical information.
Note that the recent success of the AlphaFold2, as well as other methods, in CASP 14 were achieved using second order statistics.

Our network is generic and can use both first order statistics (in terms of node attributes) and second order statistics (in terms of edge attributes). Comparing better and recent results that use second order statistics requires
additional data that are proprietary to different organizations and therefore is not done in this work.
We believe that the ProteinNet data set constitutes a great leap forward as it allows scientists to compare methods on the same footings, similarly to the impact of ImageNet \citep{ImageNet} on the computer vision community.

\paragraph{Network and optimization settings}
Throughout our experiments, we use our MimNet as described in Sec. \ref{sec:buildingMIMnet} with $n_f = 128$ with $n_{Levels} = 3$ and $T = 6$. We use the Adam optimizer with an initial learning rate of 0.0001 and a batch size of 1. Our network is trained for 100 epochs and we multiply the learning rate by a factor of 0.9 every 2 epochs. Our experiments are carried on an Nvidia Titan RTX . Our code is implemented in PyTorch \cite{pytorch}

\paragraph{Comparisons} To compare our results we use two recent works. First, for the protein folding we compare to the work of \citep{AlQuraishi2019}. While the work did not get the state of the art results, it is the only recent work known to us that uses first order information only. Furthermore, the work uses the ProteinNet data which allows us to directly compare our results to the one published. Second, for protein design, we compare the work of \citep{Strokach2020} to ours. The network proposed in the paper is named ProteinSolver and it uses a graph neural network for the solution of the design problem. The ProteinSolver network obtains the state of the art results by using a sophisticated graph representation of the protein and its features. The ProteinSolver work shows a remarkable improvement over previous design work and therefore we find it as a good benchmark.


\subsection{Protein Design}
\label{sec:designExperiment}
As discussed in Sec. \ref{sec:lossDesignTask} we measure the KL divergence between the predicted and ground-truth PSSMs. This metric allows for soft-assignments of the sequence design, which is more flexible and natural than predicting one-hot labels, due to the multiple design possibilities given a structure.
Since the data already include such probability in terms of PSSM it is only natural to use such a loss.
Our results show major improvement over a recent work ProteinSolver, reported in Tab. \ref{tab:KL}. The results for ProteinSolver were obtained by evaluating the published pre-trained model on ProteinNet dataset, for CASP $7-12$. We stress here that ProteinSolver was pre-trained on larger data, sourced from the PDB, which is the same source for ProteinNet.

\begin{table*}
  \caption{KL Divergence comparison of recent Protein-Design methods. Average of FM (novel folds) and TBM (known folds) is shown. }
  \label{tab:KL}
  \centering
  \begin{tabular}{lcccccc}
    \toprule
    Model  &\makecell{CASP $7$} & \makecell{CASP $8$}  & \makecell{CASP $9$}    & \makecell{CASP $10$}  & \makecell{CASP $11$}  & \makecell{CASP $12$} \\
    \midrule
    \midrule 

ProteinSolver \citep{Strokach2020} & 1.73 & 1.61 & 1.63  & 1.5 & 1.67 & 1.62 \\

MimNet (ours) & 0.99 & 0.88 & 0.87 & 0.83 & 0.96  & 0.95   \\

    \bottomrule
  \end{tabular}
  
    \caption{dRMSD [\r{A}] comparison of recent Protein-Folding methods. Average of FM (novel folds) and TBM (known folds) is shown.}
  \label{tab:dRMSD}
  \centering
  \begin{tabular}{lcccccc}
    \toprule
    Model  &\makecell{CASP $7$} & \makecell{CASP $8$}  & \makecell{CASP $9$}    & \makecell{CASP $10$}  & \makecell{CASP $11$}  & \makecell{CASP $12$} \\
    \midrule
    \midrule 
RGN \citep{AlQuraishi2019} & 7.45 & 6.60 & 7.60  & 8.45 & 7.95 & 8.80 \\
MimNet (ours) & 4.97 & 4.88 & 5.14 & 5.31 & 5.80  & 5.37   \\

    \bottomrule
  \end{tabular}
\end{table*}

\subsection{Protein Folding}
\label{sec:foldingExperiment}
Our folding experiment uses first-order data (PSSM and a one hot encoding matrix of the sequence design), obtained from the ProteinNet dataset. We therefore compare it to the recent work of  \cite{AlQuraishi2019} which uses identical data. Our experiments suggest that the use of GCNs can significantly improve the accuracy of protein folding as we observe a healthy reduction of the dRMSD in our method. We report the results in Tab. \ref{sec:foldingExperiment}

%


\subsection{The Significance of Reversibility}
\label{sec:ablationExperiment}

One of the key contributions of our work is the introduction of reversible networks to jointly learning protein folding and design, which have not been done until now. We therefore delve on the significance of the reversibility scheme. That is, we compare the behavior of our network when trained for both directions, versus the case of optimizing it only one direction (from sequence to coordinates and vice versa). Our results are summarized in Tab. \ref{tab:CoordToDes} $-$ \ref{tab:DesToCoord}, suggesting that coupling the learning of folding and design problems can lead to better results in both folding and design, that is, one can obtain better protein design if the network can be used to fold and vice-verse. We believe that this is due to the two problems being tightly coupled, as well as the effectively doubling of the data processed by the network. Similar behaviour is obtained when using reversible architectures to solve problems such as normalized flows \cite{yang2019pointflow}.

\begin{table}
  \caption{Comparison between Coordinates to Design (C $\rightarrow$ D) and reversible learning (C $\leftrightarrow $ D)  on CASP 7-12. Results are reported in KL-divergence score. }
  \label{tab:CoordToDes}
  \centering
  \begin{tabular}{lccc}
    \toprule
    Dataset &  \makecell{C $\rightarrow$ D}   & \makecell{D $\leftrightarrow$ C}\\
    \midrule
    \midrule 
    CASP $7$ & 1.08 & 0.99\\
    CASP $8$ & 0.87 & 0.88 \\
    CASP $9$  & 0.90 & 0.87 \\
    CASP $10$  & 0.83 & 0.83 \\
    CASP $11$  & 0.98 & 0.96 \\
    CASP $12$  & 0.97 & 0.95 \\
    \bottomrule
  \end{tabular}
\end{table}

\begin{table}
  \caption{Comparison between Design to Coordinates (D $\rightarrow$ C) and reversible learning (C $\leftrightarrow $ D) on CASP 7-12. Results are reported in dRMSD [\r{A}].}
  \label{tab:DesToCoord}
  \centering
  \begin{tabular}{lccc}
    \toprule
    Dataset  &\makecell{D $\rightarrow$ C } & \makecell{D $\leftrightarrow$ C } \\
    \midrule
    \midrule 
    CASP $7$ & 5.16 & 4.97 \\
    CASP $8$ & 4.93 & 4.88   \\
    CASP $9$  & 5.07 & 5.14 \\
    CASP $10$  & 5.29 &  5.31 \\
    CASP $11$  & 6.12 & 5.80 \\
    CASP $12$  & 5.50 & 5.37\\
    \bottomrule
  \end{tabular}
\end{table}


%
%

\section{Conclusion}
\label{sec4}

In this work we have introduced a novel approach that unifies the treatment of protein folding and protein design. Our methodology is based on a combination of two recently studied techniques in deep learning. The first is a reversible architecture. Such an architecture allows us to propagate forward and backward and therefore have a network that can propagate sequence information into  coordinates information and, more importantly for the protein design, propagate backward from a structure to a sequence. The reversible architecture is coupled with a graph based neural network which is the natural way to 
describe molecular dynamics, and in particular protein folding. Our network models the pairwise interactions between different amino acids, and uses a multiscale structure to model far field interactions. 

We use the standard dRMSD metric for the loss of the folding problem. We note that for the design problem there is no unique answer, and therefore, obtaining a single result may be meaningless. Since standard approaches for the selection of a particular sequence yield the probability of a particular residue in each location, we use the KL divergence of the PSSM as a metric. This yields a probabilistic view of the protein design problem.

We have performed extensive numerical experiments that compares both folding and design on the CASP 7-12 data sets. These data sets contain tens of thousands of proteins that we trained both on folding and design tasks. We compared the results of the protein folding to a recent work that uses only first order information. We have shown that our network performs on par or better than such networks for the folding task. However, more importantly, we have shown a significant improvement on the protein design task, achieving a KL divergence loss that is less than half of a recently published work. We attributed this success for the use of recent protein folding architectures as well as using extensive data sets that allow for better training of the proposed architecture.


\section*{Acknowledgements and Funding}
The work was funded by Genomica.ai and MITACS. ME is supported by Kreitman high-tech scholarship.

\bibliography{bib4CB2021.bib}
\bibliographystyle{icml2021}

\end{document}

%% file: math_commands.tex
\usepackage{amsmath,amsfonts,bm}

\def\eqref#1{equation~\ref{#1}}

\def\1{\bm{1}}

\DeclareMathAlphabet{\mathsfit}{\encodingdefault}{\sfdefault}{m}{sl}
\SetMathAlphabet{\mathsfit}{bold}{\encodingdefault}{\sfdefault}{bx}{n}

\def\gJ{{\mathcal{J}}}

\def\gS{{\mathcal{S}}}

\def\gX{{\mathcal{X}}}

\newcommand{\bfD}{{\bf D}}

\newcommand{\bfK}{{\bf K}}
\newcommand{\bfL}{{\bf L}}
\newcommand{\bfM}{{\bf M}}

\newcommand{\bfS}{{\bf S}}

\newcommand{\bfV}{{\bf V}}
\newcommand{\bfW}{{\bf W}}
\newcommand{\bfX}{{\bf X}}
\newcommand{\bfY}{{\bf Y}}

\newcommand{\bfone}{{\bf 1}}

\newcommand{\bftheta}{{\boldsymbol \theta}}



%% file: Fig1.tex
\begin{figure}
\begin{tikzpicture}[shorten >=1pt,->,draw=black!50, node distance=\layersep]
    \tikzstyle{every pin edge}=[<-,shorten <=1pt]
    \tikzstyle{neuron}=[circle,fill=black!75,minimum size=12pt,inner sep=0pt]
    \tikzstyle{input neuron}=[neuron, fill=blue!50];
    \tikzstyle{output neuron}=[neuron, fill=red!50];
    \tikzstyle{hidden neuron}=[neuron, fill=blue!50];
    \tikzstyle{annot} = [text width=4em, text centered]
    
    \foreach \name / \y in {2,...,4}
        \node[input neuron] at (0,-\y) {};
        
     \node[draw] at (0,-1) {Input};   
        
    \foreach \name / \y in {1,...,5}    
        \node[neuron] at (1,-\y) {};
        
    \node[draw] at (0.9,0) {$t=0$};
    \node[draw] at (2.1,0) {$t=t_1$};
    \node[draw] at (5,0) {$t=t_2$};

    \draw[red,thick] (1,-1) -- (1,-2);
    \draw[red,thick] (1,-2) -- (1,-3);
    \draw[red,thick] (1,-3) -- (1,-4);
    \draw[red,thick] (1,-4) -- (1,-5);
    \node[neuron] at (1,-1) {};
    \node[neuron] at (1,-2) {};
    \node[neuron] at (1,-3) {};
    \node[neuron] at (1,-4) {};
    \node[neuron] at (1,-5) {};

    \draw[red,thick] (2,-1) -- (2.5,-2);
    \draw[red,thick] (2.5,-2) -- (3.1,-2.8);
    \draw[red,thick] (3.1,-2.8) -- (2.5,-3.5);
    \draw[red,thick] (2.5,-3.5) -- (2,-4.5);
    \draw[red,thick] (2.5,-3.5) -- (2,-4.5);
    \draw[red,thick] (2.5,-2) -- (2.5,-3.5);
    \node[neuron] at (2,-1) {};
    \node[neuron] at (2.5,-2) {};
    \node[neuron] at (3.1,-2.8) {};
    \node[neuron] at (2.5,-3.5) {};
    \node[neuron] at (2,-4.5) {};
    
    \draw[red,thick] (4.4,-2) -- (5.5,-2);
    \draw[red,thick] (5.5,-2) -- (6.1,-2.8);
    \draw[red,thick] (6.1,-2.8) -- (5.5,-3.5);
    \draw[red,thick] (5.5,-3.5) -- (4.5,-3.5);
    \draw[red,thick] (5.5,-2) -- (5.5,-3.5);
    \draw[red,thick] (4.4,-2) -- (4.5,-3.5);
    \node[neuron] at (4.4,-2) {};
    \node[neuron] at (5.5,-2) {};
    \node[neuron] at (6.1,-2.8) {};
    \node[neuron] at (5.5,-3.5) {};
    \node[neuron] at (4.5,-3.5) {};

   \node[circle,fill=black,minimum size=3pt,inner sep=0pt]at (6.8,-2.8) {};
   \node[circle,fill=black,minimum size=3pt,inner sep=0pt]at (7.0,-2.8) {};
   \node[circle,fill=black,minimum size=3pt,inner sep=0pt]at (7.2,-2.8) {};

      \foreach \name / \y in {2.3,...,3.3}
        \node[input neuron] at (7.6,-\y) {};
   
 \draw[blue,very thick] (1,-5.3) .. controls (1.5,-5.8) and (3.5,-5.7) .. (4.9,-3.7);
 \draw[blue,very thick] (2,-4.7) .. controls (1.2,-5.0) and (3.5,-5.7) .. (4.9,-3.7);

\end{tikzpicture}
\caption{The architecture of MimNet with graph convolution layers. An embedding layer transforms the input into a latent space which then propagates through a GCN layer fed with the outputs of two previous layers. A graph that represents the protein structure is computed after each layer. for The final layer is then projected back to obtain residue coordinates. \label{fig1}}
\end{figure}

%% file: Fig2.tex
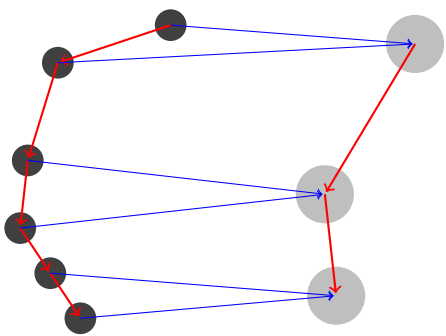
\begin{figure}
\centering
\begin{tikzpicture}[shorten >=1pt,->,draw=black!50, node distance=\layersep]
    \tikzstyle{every pin edge}=[<-,shorten <=1pt]
    \tikzstyle{neuron}=[circle,fill=black!75,minimum size=12pt,inner sep=0pt]
    \tikzstyle{tneuron}=[circle,fill=black!25,minimum size=22pt,inner sep=0pt]
    

    \node[neuron] at (5,-0) {};
    \node[neuron] at (3.5,-0.5) {};
    \node[neuron] at (3.1,-1.8) {};
    \node[neuron] at (3.0,-2.7) {};
    \node[neuron] at (3.4,-3.3) {};
    \node[neuron] at (3.8,-3.9) {};
    
    \draw[red,thick] (5,0) -- (3.5,-0.5);
    \draw[red,thick] (3.5,-0.5) -- (3.1,-1.8);
    \draw[red,thick] (3.1,-1.8) -- (3.0,-2.7);
    \draw[red,thick] (3.0,-2.7) -- (3.4,-3.3);
    \draw[red,thick] (3.4,-3.3) -- (3.8,-3.9);
    
    \node[tneuron] at (4.25+4,-0.25) {};
    \node[tneuron] at (3.05+4,-2.25) {};
    \node[tneuron] at (3.2+4,-3.6) {};
    \draw[red,thick] (4.25+4,-0.25) -- (3.05+4,-2.25);
    \draw[red,thick] (3.05+4,-2.25) -- (3.2+4,-3.6);
    
    \draw[blue] (5,0) -- (4.25+4,-0.25);
    \draw[blue] (3.5,-0.5) -- (4.25+4,-0.25);
    \draw[blue] (3.1,-1.8) -- (3.05+4,-2.25);
    \draw[blue] (3.0,-2.7) -- (3.05+4,-2.25);
    \draw[blue] (3.4,-3.3) -- (3.2+4,-3.6);
    \draw[blue] (3.8,-3.9) -- (3.2+4,-3.6);
\end{tikzpicture}
\caption{Coarsening a protein. The features of each two residues in the chain (left graph) are averaged together and a new coarse graph is computed (right) for the coarse protein. The graph Laplacian is computed directly from the rediscretized coarse protein.   \label{fig2}}
\end{figure}